\documentclass[aps,prl,showpacs,showkeys,preprintnumbers,reprint,amsmath,amssymb,superscriptaddress]{revtex4-1}

\usepackage{graphicx}
\usepackage{dcolumn}
\usepackage{bm}
\usepackage{epsf}
\usepackage{epstopdf}

\usepackage{color}

\def\duzomniejsze{<\kern-.7mm<}
\def\duzowieksze{>\kern-.7mm>}

\def\textbf#1{{\bf #1}}
\def\beq{\begin{equation}}
\def\eeq{\end{equation}}
\def\be{\begin{equation}}
\def\ee{\end{equation}}
\def\ben{\begin{eqnarray}}
\def\een{\end{eqnarray}}
\def\beqa{\begin{eqnarray}}
\def\eeqa{\end{eqnarray}}
\def\eea{\end{array}}
\def\bea{\begin{array}}
\newcommand{\bei}{\begin{itemize}}
\newcommand{\eei}{\end{itemize}}
\newcommand{\bee}{\begin{enumerate}}
\newcommand{\eee}{\end{enumerate}}

\begin{document}


\title{Can communication power of separable correlations exceed
that of entanglement resource?}
\affiliation{Faculty of Applied Physics and Mathematics,
Technical University of Gda\'nsk, 80-952 Gda\'nsk, Poland}
\affiliation{Institute of Theoretical Physics and Astrophysics, University
of Gda\'nsk, 80-952 Gda\'nsk, Poland}
\affiliation{National Quantum Information Centre of Gda\'nsk,
81-824 Sopot, Poland}

\author{Pawe\l{} Horodecki}
\affiliation{Faculty of Applied Physics and Mathematics,
Technical University of Gda\'nsk, 80-952 Gda\'nsk, Poland}

\affiliation{National Quantum Information Centre of Gda\'nsk,
81-824 Sopot, Poland}

\author{Jan Tuziemski}
\affiliation{Faculty of Applied Physics and Mathematics,
Technical University of Gda\'nsk, 80-952 Gda\'nsk, Poland}

\affiliation{National Quantum Information Centre of Gda\'nsk,
81-824 Sopot, Poland}

\author{Pawe\l{} Mazurek}
\affiliation{Institute of Theoretical Physics and Astrophysics, University
of Gda\'nsk, 80-952 Gda\'nsk, Poland}
\affiliation{National Quantum Information Centre of Gda\'nsk,
81-824 Sopot, Poland}
\author{Ryszard Horodecki}
\affiliation{Institute of Theoretical Physics and Astrophysics, University
of Gda\'nsk, 80-952 Gda\'nsk, Poland}
\affiliation{National Quantum Information Centre of Gda\'nsk,
81-824 Sopot, Poland}

\date{\today}

\begin{abstract}
The scenario of remote state preparation with shared correlated quantum state 
and one bit of forward communication [B. Daki\'c {\it et al.} Nature Physics {\bf 8}, 666 (2012)]
is considered. Optimisation of the transmission efficiency is extended to include general encoding and decoding strategies.  The importance of use of linear fidelity is recognized. It is shown that separable states cannot exceed the efficiency of entangled states by means of "local operations plus classical communication" actions limited to 1 bit of forward communication. It is proven however that such a surprising 
phenomena may naturally occur when the decoding agent has limited resources
in the sense that either (i) has to use decoding which is insensitive 
to change of coordinate system in the 
plane being in question (which is the natural choice if the receive does not know the latter)
or (ii) is forced to use bistochastic operations which may be imposed by physically 
inconvenient local thermodynamical conditions.
\end{abstract}

\maketitle

{\it Introduction .-}
It was recognized that quantum correlations provide a resource for
special tasks such as computing \cite{S1994}, teleportation \cite{BBCJPW1993}, dense coding \cite{BW1992}.
Originally the quantum advantage in realization of those task was
due to quantum entanglement. However there is a more general phenomenon
of quantum correlations involving the correlations beyond entanglement (see \cite{M2012} and references therein).
It  turned out that efficiency of the protocols involving the latter may also
exceed the efficiency of any classical solution of Deutsch-Jozsa problem \cite{KMR2006}, \cite{BBKM2006} or
Knill-Laflamme scheme \cite{DSC2008}. Quite
recently two proposals of application of quantum correlations
beyond entanglement (QCBE) have been provided. One of them have
theoretically and experimentally supported the significance of
their role in the analogue of quantum dense coding \cite{RSP1} on the level of continuous
variables. The other \cite{RSP} has addressed
the issue of the importance of QCBE for the remote state preparation (RSP).
Since RSP is one of the significant building blocks in quantum communication
the question is very important. It has already been adapted to
weak entanglement scenarios including bound entanglement \cite{MV96}
and preliminary results concerning advantage of QCBE in specific variant of RSP have been obtained \cite{CM2011}. The paper \cite{RSP} announces the  surprising possibility of
the fact that in some cases the communication power of QCBE represented by
separable states may exceed that of some entangled ones.
The authors have also provided the direct connection of
the transfer fidelity they have chosen to measure
(called geometrical discord) of quantum correlations of the resource state.
However this conclusion - unlike the experimental results of the paper fully transparent 
and impressive - seems to be not fully justified.

First of all the (quadratic) measure of transfer fidelity may be highly misleading. For instance, any classical deterministic strategy of  Bob preparing {\it completely randomly} the pure qubit state on the preagreed circle exceeds efficiencies of the originally proposed schemes in \cite{RSP} (see \cite{SM}). This makes the presence of shared quantum state irrelevant not speaking about need of 
classical communication resource.

Moreover,  only the specific class of  protocols are considered, involving von Neumann measurements (unitary operations) on Alice (Bob) side. In fact it was suggested in \cite{Adesso} that the conclusion presented in \cite{RSP} may be caused by the usage of non-optimised protocol on Bob's side.

This shows that the correct figure of merit of the state transfer in RSP protocol should be the standard linear one (see \cite{Popescu}).

Before detailed analysis it is always important to make consistent assumptions about the 
resources. In the standard LOCC paradigm the local Bloch coordinates (reference frame) are assumed 
to be known and the observers are allowed to use that knowledge 
(e.g. as for the teleportation).
However here, unlike in the teleportation scheme, the state is known to Alice, so 
to prevent direct classical transmission of its description one should put the 
restriction on the classical channel. The natural choice - compatible with the original 
scheme of RSP - is to allow for one bit of classical communication from 
sender to the receiver. So the above LOCC scheme with (i) known local Bloch coordinates 
and (ii) 1 bit of forward (form Alice to Bob) communication allowed
we shall call here {\it 1-way LOCC with 1 classical bit of information} and denote $LOCC^{\rightarrow, 1}$ (see \cite{SM}).

As we shall prove in this paper, for linear fidelity and fully optimised encoding and decoding strategies,
working under assumption of $LOCC^{\rightarrow, 1}$, there is no chance for separable state to beat 
entanglement efficiency as a resource in any RSP protocol.
Then the basic question arises if there is any other natural scenario in which the above statement does not hold. The result of \cite{RSP} give the strong evidence that it may be so in cases, when decoding is insensitive to change of coordinates in the sender's plane. However it is based on misleading fidelity and, for instance, not naturally restricted use of decoding and encoding schemes. 


Here we state the problem in a natural perspective. We use the correct fidelity and 
consider the other classes of the protocols ie. the one 
in which the Bob action should be invariant under the rotation in the plane 
form which the qubit state comes from and the other when he is forced to use  bistochastic operations. The first class is a very natural choice if Bob has no way to infer the Bloch reference frame of Alice and it will be called invariant decoding ($\Gamma_{invariant}$). The second restriction may be caused by infinite temperature of Bob's working environment ($\Gamma_{bistochastic}$). We prove by explicit analysis that both in above cases the QCBE may work better. Moreover, we find that final protocol, which was found to be optimal
under the quadratic fidelity and narrow class of protocols \cite{RSP}, happens to stay optimal (see \cite{SM})
in the two scenarios under correct, linear fidelity.

Note, that in the second scenario the final linear fidelity
for Bell diagonal states depends on the same parameters as
geometrical discord. In particular our results show
that to some extent the intuition behind the paper \cite{RSP} was correct.

\begin{figure}[h]
	\centering
		\includegraphics[width=75mm]{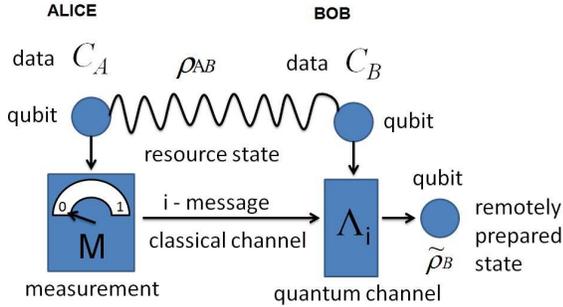}
	\label{fig1}
	\caption{ 
General scheme of RSP. For structure of initial data see Fig 2. The measurement M
is a quantum POVM and $\Lambda_i$ is  the decoding channel (see main text for detail of parametrisation). The decoding channels ${\Lambda_i}$ are supposed to belong to a fixed class $\Gamma$. In this paper we consider the most general class of all quantum channels $\Gamma_{general}$ and restricted classes: channels invariant in the $\hat{s}$ plane $\Gamma_{invariant}$ (for details see text) and the bistochastic channels $\Gamma_{bistochastic}$.
}
\end{figure}

{\it Most general RSP protocol with one bit of forward communication.-} The protocol aims to prepare at Bob's side as precisely as possible the state with 
a Bloch vector $\hat{s}$ (perpendicular to a given unit vector $\hat{\beta}$) which is known 
to Alice and not known to Bob. They share bipartite state $\rho_{AB}$ and Alice is 
allowed to send only one bit to Bob. The transmission fidelity is averaged over the unit circle constituted by all vectors on the plane perpendicular to $\hat{\beta}$. In the present analysis 
Alice is allowed to perform any generalised quantum measurement (so called POVM) 
while Bob is authorised to apply any general quantum operation represented by quantum channel. The Alice and Bob initial two-qubit state is 
\begin{equation}
\rho=\rho(\vec{x},\vec{y};T)=\frac{1}{4}[I \otimes I + \vec{x}\vec{\sigma} \otimes I + I\otimes \vec{y}\vec{\sigma} + \sum_{ij} T_{ij}
\sigma_i \otimes \sigma_j].
\label{2-qubit}
\end{equation}
The most general form of Alice binary POVM must be
a function of the following  family of parameters ${\cal A}=\{ \vec{a}, a_{+}, a_{-}\}$
and is defined by the formula  
\begin{math}
M_{\pm}=a_{\pm}I \pm \vec{a}\vec{\sigma}
\end{math}
with the probability-like parameters $a_{\pm}$ 
and vector $\vec{a}$ satisfying the conditions
\begin{eqnarray}
&& a_{+} + a_{-}=1, \ 0\leq a_{\pm}\leq 1, \nonumber \\
&& || \vec{a}||\leq min[ a_{+},a_{-}]\leq \frac{1}{2},
\label{conditions}
\end{eqnarray}
where in general both $a_{\pm}$ and $\vec{a}$ are functions of 
the unit vector $\hat{s}$ perpendicular to $\hat{\beta}$ which  has a fixed orientation during the protocol.
Finally the payoff function of the protocol is minimised over $\hat{\beta}$.

The resulting probabilities of the Alice outcomes on the state (\ref{2-qubit}) 
and the resulting states $\rho_{\pm}$ on Bob side are defined by the relations
\begin{eqnarray}
&&p_{\pm}\equiv Tr_{AB}[M_{\pm} \otimes I \rho]=(a_{\pm} \pm \vec{a} \vec{x}), \\ \nonumber
&& p_{\pm}\rho_{\pm} \equiv Tr_{A}[M_{\pm} \otimes I \rho]= \nonumber \\  \nonumber
&&=\frac{1}{2}[(a_{\pm} \pm \vec{a}\vec{x})I + (\pm T\vec{a} + a_{\pm}\vec{y})\vec{\sigma}].
\label{stany-po}
\end{eqnarray}

Bob is allowed to perform channels $\Lambda_{\pm}$ which depend upon the result $\pm$ of the Alice measurement 
and act on any qubit state 
$\rho(\vec{u})=\frac{1}{2}(I + \vec{u}\vec{\sigma})$
as
\begin{eqnarray}
&&
\Lambda_{\pm}[\rho(\vec{u})] =\frac{1}{2}[I + (T_{\pm}\vec{u} + \vec{v}_{\pm})\vec{\sigma}].\label{channel}
\end{eqnarray}

Note that by standard convexity arguments the decoding channels may always be chosen to be extremal. 

After the action of $\Lambda_{\pm}$ the final Bob state is 
\begin{equation}
\tilde{\rho}_{B}=\sum_{r=\pm}p_{r}\Lambda_{r}(\rho_{B}^{(r)})=\frac{1}{2}(I + \vec{r}\vec{\sigma}),
\end{equation} 
with the final Bob Bloch vector 
\begin{equation}
\vec{r}=\sum_{r=\pm}T_{r}(a_{r}\vec{y} + r T\vec{a}) + (a_{r}+r\vec{a}\vec{x}) \vec{v}_{r}].
\end{equation}

{\it Probabilistic fidelity .-} For fixed $\hat{s}$ the {\it probabilistic fidelity} of the success in remote state 
preparation of a pure state $\rho(\hat{s})=\frac{1}{2}(I + \hat{s}\vec{\sigma})$ is defined as (see \cite{Popescu}):
\begin{math}
F(\hat{s})=\frac{1}{2}(1 +  \vec{r} \hat{s}).  
\label{fid}
\end{math}
Averaging over $\hat{s}$ gives 
\begin{math}
\bar{F}=\frac{1}{2}(1 + G),
\end{math}
 where the fidelity parameter is 
\begin{eqnarray}
&&G=G(\rho;\hat{\beta};{\cal A}, {\cal T})= \int d \hat{s} (\vec{r}\hat{s}) = \\ \nonumber && \int d\hat{s} [(T_{+} - T_{-})T\vec{a} + (\vec{v}_{+}-\vec{v}_{-})\vec{x} \vec{a} + \nonumber \\ \nonumber
&& a_{+}(T_{+}\vec{y}+\vec{v}_{+}) + a_{-}(T_{-}\vec{y}+\vec{v}_{-})]\hat{s}.
\label{average}
\end{eqnarray}
Here $\rho=\rho(\vec{x},\vec{y},T)$, $\hat{\beta}$ defines 
the plane to which the vectors $\hat{s}$ belongs and the explicit dependence on the encoding ${\cal A}=\{ \vec{a}, a_{+}, a_{-}\}$ 
and decoding strategy ${\cal T}=\{T_+,\vec{v}_{+};T_-,\vec{v}_{-}\}$ is written.
The full range of parameters describing the encoding ${\cal A}$ is 
written explicitly in (\ref{conditions}).
The  range of parameters $\cal{T}$ is determined by the structure of the extremal one-qubit channels \footnote{$T_{\pm}=\mathcal{O}^{(1)}_{\pm} T_{\pm}^{0}(\mathcal{O}^{(2)}_{\pm})^{T}, \vec{v}_{\pm}=\mathcal{O}^{(1)}_{\pm}\vec{v}_{\pm}^{0}$, where $\mathcal{O}^{(1)}_{\pm}, \mathcal{O}^{(2)}_{\pm}$ are arbitrary rotations combined with representation of completely positive trace preserving maps. Here we use a family given by $\vec{v}_{\pm}^{0}=[0,0,\sin u_{\pm}\sin w_{\pm}]$, $T_{+}^{0}=[ \cos u_{+}, \cos w_{+},\cos u_{+}\cos w_{+}]$, $T_{-}^{0}=[cos u_{-}, \cos w_{-}, \cos u_{-}\cos w_{-}]$
with $u_{\pm}\in[0,2\pi)$, $w_{\pm}\in[0,\pi)$ \cite{Ruskai}. This is the most general form of any one-qubit channel belonging to the closure of the set of extreme one-qubit channels. By simple convexity argument it is enough to consider only these channels.}.

{\it Optimal RSP protocol for separable states. Advantage of entangled states in $LOCC^{\rightarrow,1}$.- } Here we keep the assumption of $LOCC^{\rightarrow,1}$ in which Alice and Bob naturally share the reference frame on the Bloch sphere. We may choose the coordinates as $\left\{\hat{\beta} , \hat{e}, \hat{e}' \right\}$ where $\hat{\beta}\times \hat{e} = \hat{e}'$ and $\left\{\hat{e}, \hat{e}' \right\}$ represent the coordinates system in the $\hat{s}$ plane.
    
Because fidelity is convex, for separable states it is sufficient to consider pure states. For pure states $p_\pm = a_ \pm (\hat{s}) + \vec{a}(\hat{s})\hat{ \xi}$, where $\hat{\xi}$ is Alice Bloch vector. The reduced state of Bob is $\rho_\pm = \frac{1}{2} \left(I + \vec{n}_\pm \hat{\sigma}\right)$, where $\vec{n}_\pm$ is Bob Bloch vector transformed by respective channel. Then (\ref{average})  is of a form $G=\frac{1}{2} \int d\hat{s} p_+(\hat{s})\left(\vec{n}_+(\hat{s})-\vec{n}_-(\hat{s})\right)\hat{s}$. The bracket $\left(\vec{n}_+(\hat{s})-\vec{n}_-(\hat{s})\right)\hat{s}$ attains maximal value for  $\vec{n}_+ = \hat{e}, \vec{n}_- = -\hat{e} $ ($\vec{n}_+ = -\hat{e}, \vec{n}_- = \hat{e} $) such that $\hat{e} \hat{s} >0$ ($\hat{e} \hat{s} <0$), where $\hat{e}$ is an a priori known unit vector (see above). Then setting $p_+(\hat{s})=1$ is optimal. In this case Alice POVM reduces to identity and Bob prepares vector $\pm \hat{e}$ depending on the sign of $\hat{e} \hat{s}$. Note that this protocol is independent of an input state and its fidelity is
\begin{eqnarray}
\bar{F} = \frac{1}{2}\left(1 + \frac{2}{\pi} \int^{\frac{\pi}{2}}_{0} d \theta \cos \theta \right) = \frac{1}{2}\left(1 + \frac{2}{\pi}\right).
\end{eqnarray}
As a result using separable states cannot lead to better fidelity than using entangled states ($\bar{F}(\rho_{ent}) \geq \bar{F}(\rho_{sep}) $). In the case when for an entangled state there is no better strategy one can always use this protocol.    
In order to answer the question in what scenario separable states can have advantage over entangled states we will perform optimisation over Alice POVMs.

{\it Optimisation over Alice POVMs for arbitrary quantum state .-} Optimisation of the formula (\ref{average}) over the strategies ${\cal A}$ can be performed as follows (see \cite{SM}): we have three sets in the unit circle on the $\hat{s}$ plane: 
$\Omega_{0}$, $\Omega_{\pm}$
defined as 
\begin{math}
\label{sets}
\Omega_{0}= \{ \hat{s}: \hat{s}\hat{\beta}=0, ||M^{T}\hat{s}||\geq |(\vec{V}_{+}-\vec{V}_{-})\hat{s}|\},\Omega_{+}=\{\hat{s}: \hat{s}\hat{\beta}=0, ||M^{T}\hat{s}|| < (\vec{V}_{+}-\vec{V}_{-})\hat{s}\},  \Omega_{-}=\{\hat{s}: \hat{s}\hat{\beta}=0, ||M^{T}\hat{s}|| < (\vec{V}_{-}-\vec{V}_{+})\hat{s}\},
\end{math}
where $M = (T_+ - T_-)T + (\left|\vec{v}_+\right\rangle-\left|\vec{v}_-\right\rangle)\left\langle \vec{x}\right|$, $\vec{V}_+=(T_+ \vec{y} + \vec{v}_+)$, $\vec{V}_-=(T_-\vec{y} + \vec{v}_-)$. Let us define $\Omega_{0}^{+}=-\Omega_{0}^{-}$ as any of two subsets of original $\Omega_0$ such 
that $\Omega_0=\Omega_{0}^{+} \cup \Omega_{0}^{-}$. 
The final formula optimised over ${\cal A}$ is of the form
\begin{eqnarray}
&& max_{{\cal A}} G(\rho;\hat{\beta};{\cal A},{\cal T})= \nonumber \\
&& \int_{\Omega_{0}^{+}} d \hat{s} ||M^{T}\hat{s}|| + \int_{\Omega_{+}} d \hat{s}  (\vec{V}_{+} - \vec{V}_{-})\hat{s}
\label{integral-kcso}.
\end{eqnarray}

{\it Optimisation in the case of $\Gamma_{invariant}$.-} In this section we will investigate a case of the protocol, in which
Bob's strategy is independent of the setting on Alice side. This
means that Alice, after establishing decoding strategy with Bob, can e.g.
change type of input state by choosing different angle $\varphi$ or her
coordinates system in the plane orthogonal to $\hat{\beta}$ and Bob's
strategy should remain optimal. As a consequence, this strategy
cannot depend on the parametrisation of the input Bloch vector.
Technically in this case decoding operations should be restricted to the
class, which is invariant under averaging in the plane orthogonal to
$\hat{\beta}$ or always look the same after any rotation in that plane.
We will denote this class as $\Gamma_{invariant}$.  For operations belonging to $\Gamma_{invariant}$  we have that ${\cal T} \left(\left\{T_ \pm \right\},  \left\{v_ \pm\right\}\right) = {\cal \tilde{T}} \left(\left\{\tilde{T}_ \pm \right\},  \left\{\tilde{v}_ \pm\right\} \right) $, where      
\begin{math}
 \tilde{T}_{\pm} =\frac{1}{2 \pi} \int^{2 \pi}_{0} d \varphi O_{\hat{s}} (\varphi) T_{\pm} O^T_{\hat{s}}(\varphi),
 \tilde{\vec{v}}_{\pm} = \frac{1}{2 \pi} \int^{2 \pi}_{0} d \varphi O_{\hat{s}} (\varphi) \vec{v}_{\pm},
\end{math}  
where $O_{\hat{s}}(\varphi)$ denotes rotation in the $\hat{s}$ plane. As a result of averaging $\tilde{T}_{\pm} = \text{diag} [t_\pm, \tilde{T}^{(1)}_\pm]$, where $\left|t_\pm\right| \leq 1, \left\|\tilde{T}^{(1)}_\pm\right\| \leq 1$ and $\tilde{T}^{(1)}_\pm$ are 2x2 matrices acting in the $\hat{s}$ plane an invariant under any rotation $\tilde{T}^{(1)}_\pm = O \tilde{T}^{(1)}_\pm O^T $. The use of this class is natural also from game-like perspective: let us allow Bob to use arbitrary channel ${\cal T}$. Since he does not know the coordinates he must average his decoding strategy over all possible orientations of reference frame on the $\hat{s}$ plane. This results in decoding from $\Gamma_{invariant}$. As a consequence we get that $\tilde{\vec{v}}_{\pm}$ have no components parallel to $\hat{s}:$
\begin{eqnarray}
&&\vec{r}\hat{s}=\sum_{r=\pm}[\tilde{T}^{(1)}_{r}(a_{r}\vec{y} + r T\vec{a})]\hat{s}.
\end{eqnarray}  
By setting $M=(\tilde{T}^{(1)}_+-\tilde{T}^{(1)}_-)T$ and $\vec{V}_{\pm}=\tilde{T}^{(1)}_{\pm} \vec{y}$ and inserting it into (\ref{integral-kcso}), we obtain optimised formula for $max_{{\cal A}} G(\rho(\vec{x},\vec{y},T);\hat{\beta};{\cal A},{\cal \tilde{T}})$. In the case of an isotropic correlations $T= - \lambda I$ we have following facts:

{\it Fact 1.- } We can always decompose $\vec{y}$ as $\vec{y}= \left\|\hat{y}\right\| \left[\alpha \hat{u} + (1-\alpha) \hat{\beta} \right]$. Then  the 
formula $max_{{\cal A}} G(\rho(\vec{x},\vec{y},-\lambda I);\hat{\beta};{\cal A},{\cal \tilde{T}})$ is monotonic function of parameter $\alpha = \left| \vec{y} \hat{u} \right|$.

{\it Fact 2.- } The optimisation over the $\Gamma_{invariant}$ class (which naturally corresponds to the situation with an  unknown coordinates
system, as in case b) of Fig. 2) yields $min_{ \hat{\beta}} max_{{\cal A},\tilde{{\cal T}}} G(\rho(\vec{x},\vec{y},-\lambda I);\hat{\beta};{\cal A},\tilde{{\cal T}})  = \lambda$ and then consequently
\begin{eqnarray}
\bar{F}_{invariant}(\rho(\vec{x},\vec{y},-\lambda I) )= \frac{1}{2}\left(1+ \lambda \right).
\end{eqnarray}
For details of the proofs of the above see \cite{SM}. Now following \cite{RSP}
consider the following class: $\rho(t\hat{z},t\hat{z},-\lambda I)$; or in other words the states with the parameters: $T=-\lambda I$, $\vec{x}=\vec{y}=t\hat{z}$ where the positivity condition determines the following range of parameter $t$:
$|t|\leq \frac{1- \lambda}{2}$. For any fixed nonzero  $\lambda$ there are entangled states in that 
class namely the ones satisfying in addition the inequality $|t| > \frac{1}{2} \sqrt{1-2 \lambda -3 \lambda^2}$. 
All of those entangled states $\rho(\vec{x},\vec{y},-\lambda I)$ with $\lambda < \frac{1}{3}$ will -  due to the Fact 2 - have worse RSP fidelity (under the restriction of  unknown coordinate system)  than the separable states $\rho(\vec{0},\vec{0},-\lambda' I)$ 
with $\lambda' \in (\lambda,\frac{1}{3})$ .  (This comprises as special cases considered in Ref. \cite{RSP}: the 
separable case $\lambda'=\frac{1}{3}$, $t=0$ and an entangled one with  $\lambda=\frac{1}{5}$, $t=\frac{2}{5}$).  The overall conclusion is that whenever Bob does not know the coordinates of Alice in the $\hat{s}$ plane then entanglement may be less useful than quantum correlations beyond it, i.e. the ones contained in separable states.

\begin{figure}[h]
	\centering
		\includegraphics[width=80mm]{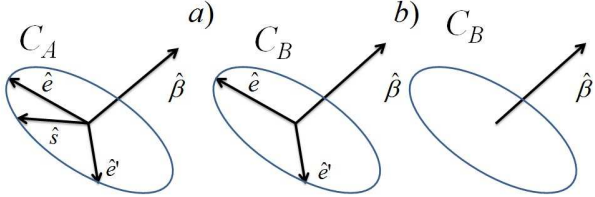}
	\label{fig2}
	\caption{Initial data for Alice and Bob with a) or without b) shared reference frame on the $\hat{s}$ plane.}
\end{figure}

{\it Optimal RSP in the case of bistochastic channels for Bell diagonal states  .-} Let us consider a situation when temperature of Bob's environment is infinite. Then he is restricted to use bistochastic channels.
Detailed analysis shows that for Bell diagonal states the formula (\ref{integral-kcso}) 
can be fully optimised. Here by Bell diagonal states we mean 
all the states that are local unitary (i.e. $U_{1} \otimes U_{2}$ type) rotations
of the states diagonal in the standard Bell basis $\Psi_{\pm}=\frac{1}{\sqrt{2}} (|0 0\rangle \pm  
|1 1\rangle)$ and $\Phi_{\pm}=\frac{1}{\sqrt{2}} (|0 1\rangle \pm  
|1 0\rangle)$. It is known that \cite{MRH-96} all such states can 
be represented by $\rho(\vec{0},\vec{0},T)$. Then after some 
algebra (see \cite{SM}) one gets $\min_{\beta} max_{{\cal A}, {\cal \check{T}}} G(\rho(\vec{0},\vec{0},T);\hat{\beta};{\cal A},{\cal \check{T}}) =\frac{2 |t_2|}{\pi} E \left(\sqrt{1-\frac{t^2_1}{t^2_2}}\right)$ and
\begin{eqnarray}
\label{resbd}
\bar{F}_{bistochastic} = \frac{1}{2}\left[1 + \frac{2 |t_2|}{\pi} E \left(\sqrt{1-\frac{t^2_1}{t^2_2}}\right)\right],
\end{eqnarray}
where $t^2_1, t^2_2$ are two lowest eigenvalues of $T^T T$, E(x) is complete elliptic integral of the second kind \cite{NIST} and ${\cal \check{T}}$ denotes Bob bitochastic decoding. In this case it is also possible to show that there exists separable states leading to higher fidelity of RSP protocol than entangled states. Let us consider two Bell diagonal states with following correlation tensors: $T_1 =\left[ - \frac{1}{3}, - \frac{1}{3}, - \frac{1}{3}\right], \; T_2 =\left[ - \frac{1}{3} - 2 \epsilon, - \frac{1}{3} +\frac{\epsilon}{2}, - \frac{1}{3} +\frac{\epsilon}{2} \right] $ with $\epsilon > 0$. The set of separable Bell diagonal states is specified by condition $\left|t_1\right| + \left|t_2\right| + \left|t_3\right| \leq 1$. Clearly the state corresponding to $T_1$ is separable whereas that corresponding to $T_2$ is not. Using (\ref{resbd}) one immediately obtains that 
$\bar{F}_{bistochastic}(\rho(\vec{0},\vec{0},T_1))=\frac{2}{3} > \frac{2}{3}-\frac{\epsilon}{4}=\bar{F}_{bistochastic}(\rho(\vec{0},\vec{0},T_2))$.

Interestingly (\ref{resbd}) depends only on the two smallest eigenvalues of $TT^T$. Since the geometric discord is in this case of the form $D(\rho(\vec{0},\vec{0},T))=\frac{1}{4}(t^2_1+t^2_2)$ the optimised fidelity
 depends on the same parameters like the one used in \cite{RSP}.
This shows that in the case of bistochastic decoding the presented result based on standard fidelity and the one based on quadratic fidelity are consistent.
  
{\it Conclusions .-}   It is known that quantum correlations without entanglement, contained in separable states may be useful in quantum
information processing. The basic issue is whether they may outperform entanglement in any case. 
Our analysis shown that one should be careful in comparison of the two resources.  In fact two-qubit separable states {\it  cannot} outperform two-qubit entanglement in the process of  remote state preparation of quantum bit under most general assumptions ie. LOCC protocol natural for the problem - the one with classical communication restricted to one bit of forward communication.  This lies in the heart of the balance of  quantum resources within so called LOCC scenario: whenever initial entanglement is too weak, 
 Bob may remove entanglement and prepare the optimal state on his own still achieving the best  efficiency 
provided by all  separable correlations based protocols. Thus any protocol with initial entanglement 
cannot be worse than the one with separable state.

   The apparent contradictions to the above may only take place if one use nonstandard figure of merit. 
To be correct the latter must make a difference between the orthogonal states in any basis and 
the standard linear fidelity works well from that perspective.

   However it turns out that all the above does not prevent quantum information from ,,quantum separability advantage'' 
in the cases when the users have extra restrictions, somehow very natural, on the operations they are allowed to apply. Here we have shown that if in the RSP protocol of a qubit state with two-qubit correlations the
receiver is restricted to the decoding class, which reflects
his ignorance about the coordinates in the plane the origin
state comes from, then separability can work better
than entanglement. The second scenario when the latter may happen is the one when the receiver is forced to use bistochastic 
decodings. Then, whenever Alice and Bob share Bell diagonal  quantum state, the linear fidelity
of the protocol depends on the same set of parameters as geometric quantum discord.

    The latter result is even more intriguing, when one realises that the restriction of bistochastic character of the 
decoding may be interpreted as a presence of ,,thermodynamically unsuitable'' anciallas 
namely those of infinite temperature. Note that in this case quantum correlations beyond entanglement 
help better than entanglement itself and that one option to discriminate those correlations 
form classical ones is just the thermodynamical picture of local engines (see \cite{Oppenheim,Zurek}).
 This suggests that possible thermodynamical perspective of the discussed protocol (and also practical aspects 
of  other protocols aimed in using quantum correlations beyond entanglement)  should be examined more 
 in future.

{\it Acknowledgements.-}  PH and RH thank A. Miranowicz and  R. Chhajlany for discussions.
This paper is supported by Polish Ministry of Science and Higher Education Grant
no. IdP2011 000361.  PM is supported by the Foundation for Polish Science International PhD Projects Programme
co-financed by the EU European Regional Development Fund. JT is supported by EU Project QESSENCE.

{\it Supplemental Material .-}

The structure of Supplemental Material is as follows: firstly we restate the scenario considered in \cite{RSP} and support our claim that using quadratic figure of merit is misleading because any classical deterministic strategy of Bob preparing pure state along {\it arbitrary} fixed axis exceeds efficiencies of the originally proposed protocol in \cite{RSP}. We calculate the fidelity of such protocol. Then firstly we introduce definitions of possible LOCC with restricted amount of communication. Subsequently we provide detailed derivation of optimised fidelity over Alice POVMs for arbitrary quantum state as well as optimisation in the case of $\Gamma_{invariant}$ for isotropic correlations and in the case of biostatistic channels for Bell-diagonal states.

{\it The protocol of Ref. \cite{RSP}  and the fidelity issue.-}
In Ref. \cite{RSP} the authors consider the specific subclass of
the protocols from Fig. 1 of the main text, ie. they only allow
von Neumann measurement $M$ and very specific unitary decodings 
($\Lambda_+=I$, $\Lambda_-=-I$) that correspond to identity or reflection on the considered circle on the Bloch sphere.
They also exploit the quadratic transfer fidelity:
\begin{equation}
{\cal P}=\int d\hat{s} ({\hat s}\vec{r})^{2}
\label{fidelity}
\end{equation}
Here the vector $\hat s $ corresponds to the Bloch vector of the qubit
state of Alice $\rho({\hat s}) \frac{1}{2}(I + {\hat s}\sigma)$
that is to be prepared on Bob side.
Its description is  included in the Alice data $C_{A}$ depicted on Fig. 1,
and explained in Fig. 2 of the main text. The central assumption is
that the Bloch vector of Alice qubit is chosen randomly from
the circle, location of which is determined by
some unit vector $\hat \beta$
(since the circle is located on the plane perpendicular to $\hat \beta$).
While the Ref. \cite{RSP} does not specify
which of the two structures of the Bob's data of Fig. 2 
it considers, it seems to
tacitly assume the case b) which does not allow
Bob to know the reference frame in the considered plane.
After the protocol Bob gets the final state
$\tilde{\rho}_{B}$ (see Fig. 1 of the main text) and
its Bloch vector is put into the formula
(\ref{fidelity}) above.

For given shared state $\rho_{AB}$
the fidelity (\ref{fidelity}) is first maximised over
all allowed protocols (which are
(i) Alice von Neumann measurement plus (ii) Bob specific unitary
encoding) and then minimised with respect to the orientation
of the vector $\hat \beta$ which gives for an initial state
$\rho_{AB}$ the optimal quadratic fidelity ${\cal P}_{opt}$.

The authors find quantumly correlated (in a standard sense,
reported by nonzero geometric discord in their paper) but non-entangled Werner
state
 $\rho_{AB}=\rho(0,0, - \lambda' I)$
(here $I$ is the 3x3 identity matrix, see the formula (1))
with $\lambda'=\frac{1}{3}$ for which ${\cal P}_{opt}=\frac{1}{25}$
is strictly larger than its value ${\cal P}_{opt}=\frac{1}{9}$
offered by the following entangled state $\rho_{AB}=\rho(t\hat{z},t\hat z,
- \lambda I)$
with $t=\frac{1}{5}$ and $\lambda'=\frac{2}{5}$. 

On that basis the conclusion of Ref. \cite{RSP} is made, that quantum correlations beyond entanglement
represented by non-zero discord of the separable state above
make quantum correlations beyond entanglement better then
entanglement itself in remote state preparation.

However the incorrect choice of the (quadratic) fidelity ruins
this conclusion in the most natural, practical sense.
It turns out that - unless we specify explicitly restrictions (as done in
the main part of the present paper) - in both two cases
there exist a {\it trivial} protocol which supersedes them.
In fact given any of the  two-qubit states above
Bob may ignore the Alice message and just take randomly chosen
state $\tilde{\rho}_{B}$ with its Bloch vector randomly
located on the circle perpendicular to $\hat \beta$ and
get the fidelity $\frac{1}{2}$ which is much better than
$\frac{1}{25}$ and $\frac{1}{9}$ above.
Indeed consider a case of the protocol in which Bob, regardless of the Alice message, produces at random a pure state with a Bloch vector belonging to the $\hat{s}$ plane. Employing fidelity proposed in \cite{RSP} we obtain 
\begin{eqnarray}
&&F=\min_{\hat{\beta}} \left\langle (\hat{r} \hat{s})^2 \right\rangle = \left\langle (\hat{r} \hat{s})^2 \right\rangle =\int d \; \hat{s} (\hat{r} \hat{s})^2  = \\ \nonumber  &&\frac{1}{2 \pi}\int^{2 \pi}_{0} d \; \varphi \cos^2 \varphi = \frac{1}{2},
\end{eqnarray} where we used the invariance of the measure on the unit circle. Because this fidelity is higher than these considered in \cite{RSP} ($\frac{1}{9}$ and $\frac{1}{25}$ for separable and entangled state respectively), it may seem that the random protocol is better choice than more sophisticated strategies. However, this is only because of misleading choice of protocol fidelity.

{\it LOCC with restricted amount of communication .-}
Here we introduce formal definitions of possible subclasses of LOCC with certain amount of restricted communication in each round

{\it Class of "one-way" forward LOCC operations with restricted amount of communication .-}  Using this class of operations Alice is allowed to send only certain number of bits to Bob, eg. in the case of RSP protocol only one bit can be sent. These operations belong class {\it C2a} described in \cite{RMP}. We will denote operation belonging to this class as $LOCC^{\rightarrow, c}$, where c is fixed number of forward classical bits of communication.  

{\it Class of "one-way" backward LOCC operations with restricted amount of communication .-}  The situation is the same as in the above class but with the roles of Alice and Bob interchanged. These operations belong class {\it C2b} described in \cite{RMP}. In analogy to the previous case, we will denote this class as $LOCC^{\leftarrow, c}$.  

{\it Class of class of "two-way" operations with restricted amount of communication .-} In this class Alice and Bob are allowed to send only certain number of bits to each other. Note that in general numbers do not have to be equal ie. Alice can be allowed to send different number of bits than Bob. These operations belong class {\it C3} described in \cite{RMP}. In analogy to the previous cases, we will denote this class as $LOCC^{\leftrightarrow, c}$. Note that
here the index $c$ represents the sequence of numbers that put the limits on the communicated bits in each round
of the protocol since one allows many rounds of communication.

{\it Optimisation over Alice POVMs .-}
Here we shall optimise the quantity 
\begin{eqnarray}
\nonumber && G = \int d\hat{s} [(T_{+} - T_{-})T\vec{a} + (\vec{v}_{+}-\vec{v}_{-})\vec{x} \vec{a} + \nonumber \\
&& a_{+}(T_{+}\vec{y}+\vec{v}_{+}) + a_{-}(T_{-}\vec{y}+\vec{v}_{-})]\hat{s} \nonumber
\end{eqnarray}
with respect to  ${\cal A}$ (to be concise we shall omit all the arguments in its notation).
Define the matrix $M = (T_+ - T_-)T + (\left|\vec{v}_+\right\rangle-\left|\vec{v}_-\right\rangle)\left\langle \vec{x}\right|$, and $\vec{V}_+=(T_+ \vec{y} + \vec{v}_+)$, $\vec{V}_-=(T_-\vec{y} + \vec{v}_-)$. Then the above function is of the form
\begin{eqnarray}
G=\nonumber \int d\hat{s} [M \vec{a} + a_{+}\vec{V}_{+} + a_{-}\vec{V}_{-}]\hat{s},
\label{integral-1}
\end{eqnarray}
where the vector $\vec{a}$ and the scalars $a_{\pm}$ depend in general on $\hat{s}$ and satisfy the conditions (2) (main text).
 Let us put $\vec{a}=a\hat{a}$, where $0 \leq a=||\vec{a}||\leq a_{\pm}$. Clearly the best 
choice to maximise the value of $G$ is to put $\hat{a}$  parallel to the vector $M^{T}\hat{s}$ 
or, in other words, $\hat{a}=\frac{M^{T}\hat{s}}{||M^{T}\hat{s}||}$. Then 
the value of the integral becomes 
$G =\int d \hat{s} a[||M^{T}\hat{s}|| + a_{+}\vec{V}_{+} + a_{-}\vec{V}_{-}]\hat{s}$, which 
may be further optimised with respect to $a$ by taking its maximal allowed value $a=min[a_+,a_-]$.

Eventually, this gives the function optimised over $\vec{a}$ for fixed $a_{\pm}$ and all the other parameters:
\begin{equation}
G=\nonumber \int d\hat{s} ||[M^{T}\hat{s}||min[a_+,a_-] + a_{+}\vec{V}_{+} + a_{-}\vec{V}_{-}]\hat{s}.
\label{integral-2}
\end{equation}

Using notation $M'=||[M^{T}\hat{s}||\geq 0$, $A_{\pm}=\vec{V}_{\pm}\hat{s}$,
we may carefully consider the maximum of 
\begin{equation}
f(p)= M' min[p,1-p] + p A_{+} +(1-p)A_{-} 
\end{equation}
over the interval $p\in [0,1]$, where we put $p=a_+$ and $1-p=a_{-}$ for conciseness.
The above function has the following maxima: 

(a) if $M'\geq |A_{+} - A_{-}|$, then $max_{p\in[0,1]} f(p)= \frac{M+A_{+}+A_{-}}{2}$ achieved at $p=\frac{1}{2}$;

(b) if $M' < |A_{+} - A_{-}|$, then either (i) $max_{p \in[0,1]} f(p)=A_{+}$ for $A_{+} - A_{-} > 0 $ (achieved at $p=1$)
or (ii) $max_{p \in[0,1]} f(p)=A_{-}$ for $A_{-} - A_{+} > 0 $ (achieved at $p=1$).

In case (a) the strategy of Alice is naturally the one of Ref. \cite{RSP}; she performs the von Neumann measurement with the projections $P_{\pm\hat{a}}=\frac{1}{2}(I \pm \hat{a}\vec{\sigma})$.
An intriguing strategy of Alice in case (b) is that she just does {\it nothing}
(since then the POVM is the identity) and puts the message $r$ to Bob depending on the 
sign of $(A_{+} - A_{-})=(\vec{V}_{+}-\vec{V}_{-})\hat{s}$.  Quite remarkably this strategy gives {\it always nonnegative} contribution 
form the part of the integral (\ref{integral-2}) involving the vectors $\vec{V}_{\pm}$.
 
We have then the three sets in the unit circle on the $\hat{s}$ plane: 
$\Omega_{0}$, $\Omega_{\pm}$
defined as 

(i) $\Omega_{0}= \{ \hat{s}: \hat{s}\hat{\beta}=0, ||M^{T}\hat{s}||\geq (\vec{V}_{+}-\vec{V}_{-})\hat{s}\};$
(ii) $\Omega_{+}=\{\hat{s}: \hat{s}\hat{\beta}=0, ||M^{T}\hat{s}|| < (\vec{V}_{+}-\vec{V}_{-})\hat{s}\} ;$
(iii) $\Omega_{-}=\{\hat{s}: \hat{s}\hat{\beta}=0, ||M^{T}\hat{s}|| < (\vec{V}_{-}-\vec{V}_{+})\hat{s} \}.$

The final formula optimised over ${\cal A}$ is of the form 
\begin{eqnarray}
&& max_{{\cal A}} G(\rho(\vec{0},\vec{0},T);\hat{\beta};{\cal A},{\cal T})= \nonumber \\
&& \int_{\Omega_{0}} d \hat{s} \frac{||M^{T}\hat{s}|| + \vec{V}_{+}\hat{s} + \vec{V}_{-}\hat{s}}{2} + \nonumber \\
&& \int_{\Omega_{+}} d \hat{s}  \vec{V}_{+}\hat{s}  + \int_{\Omega_{-}} d \hat{s}  \vec{V}_{-}\hat{s}. 
\label{integral-3}
\end{eqnarray}

We have also a following

{\it Observation .- }\\
i) $\Omega_0 = -\Omega_0$, $\Omega_0$ is symmetrical,
ii) $\Omega_+=-\Omega_-$, i.e. after the reflection the sets are equal. \\
From i) we get that $ \int_{\Omega_{0}} d \hat{s} \frac{||M^{T}\hat{s}|| + \vec{V}_{+}\hat{s} + \vec{V}_{-}\hat{s}}{2} = \int_{\Omega_{0}} d \hat{s} \frac{||M^{T}\hat{s}||}{2}$, and from ii)$ \int_{\Omega_{+}} d \hat{s}  \vec{V}_{+}\hat{s}  + \int_{\Omega_{-}} d \hat{s}  \vec{V}_{-}\hat{s} = \int_{\Omega_{+}} d \hat{s}  (\vec{V}_{+} - \vec{V}_{-})\hat{s}$. 
Let us define $\Omega_{0}^{+}=-\Omega_{0}^{-}$ as any of two subsets of original $\Omega_0$ such 
that $\Omega_0=\Omega_{0}^{+} \cup \Omega_{0}^{-}$. 

The final formula optimised over ${\cal A}$ is of the form
\begin{eqnarray}
&& max_{{\cal A}} G(\rho;\hat{\beta};{\cal A},{\cal T})\equiv G(\rho;\hat{\beta};{\cal A^*},{\cal T})  \nonumber \\
&& \int_{\Omega_{0}^{+}} d \hat{s} ||M^{T}\hat{s}|| + \int_{\Omega_{+}} d \hat{s}  (\vec{V}_{+} - \vec{V}_{-})\hat{s},
\end{eqnarray}
where ${\cal A^*}$ denotes optimal Alice measurement (either von Neumann or trivial one).

{\it Proof of Fact 1 .- } 
Let us consider $\alpha' > \alpha$, where $\alpha = \left| \vec{y} \hat{u} \right|$. Parameters $\alpha, \alpha'$ correspond to two different orientations of $\vec{y}$ with respect to $\hat{s}$ plane ie. $\vec{y}= \left\|\hat{y}\right\| \left[\alpha \hat{u} + (1-\alpha) \hat{\beta} \right]$and  $\vec{y'}= \left\|\hat{y}\right\| \left[\alpha' \hat{u} + (1-\alpha') \hat{\beta} \right]$ . For $\alpha'$ ($\alpha$) we will denote solutions of inequalities defining the sets as $\Omega^{+'}_0$, $\Omega^{'}_+$, ($\Omega^+_0$, $\Omega_+$), similarly  $||M^{T}\hat{s}|| = f'$ $(\vec{V'}_{+} - \vec{V'}_{-})\hat{s} = g'$ ($||M^{T}\hat{s}|| = f$ $(\vec{V}_{+} - \vec{V}_{-})\hat{s} = g$). Let us recall that here we consider only
the restricted class of the invariant (equivalently averaged) decodings ${\cal\tilde{T}}$. It follows from the definition of the sets that $\Omega^{+'}_0 < \Omega^{+}_0$ and $\Omega^{'}_+ > \Omega_+$ as well as $f'=f=\lambda||(\tilde{T}^{(1)}_+-\tilde{T}^{(1)}_-)^{T}\hat{s}||
$. We can rewrite $g'$ as $g' = (\tilde{T}^{(1)}_+-\tilde{T}^{(1)}_-)\vec{y'}\hat{s} = \vec{y'} (\tilde{T}^{(1)}_+-\tilde{T}^{(1)}_-)^T \hat{s} = \vec{y}' \vec{w}(\hat{s}) $ and as a consequence the following relation holds  $g'=  \vec{y'} \vec{w}(\hat{s}) = \left\|\vec{y'}\right\| \alpha ' \hat{u} \vec{w}(\hat{s})  = \left\|\vec{y}\right\| \alpha ' \hat{u} \vec{w}(\hat{s}) > \left\|\vec{y}\right\| \alpha  \hat{u} \vec{w}(\hat{s}) = g$ . Thus we can write
\begin{eqnarray}
&&\max_{{\cal A}} G(\rho;\hat{\beta'};{\cal A},{\cal \tilde{T}}) =\int_{\Omega_{0}^{+'}} d \hat{s} f' + \int_{\Omega^{'}_{+}} d \hat{s}  g' =  \\ \nonumber &&\int_{\Omega_{0}^{+'}} d \hat{s} f + \int_{\Omega^{'}_{+} \setminus \Omega_{+} } d \hat{s} g' +  \int_{ \Omega_{+} } d \hat{s} g' \geq \\ \nonumber &&\int_{\Omega_{0}^{+'}} d \hat{s} f + \int_{\Omega^{'}_{+} \setminus \Omega_{+} } d \hat{s} f +  \int_{ \Omega_{+} } d \hat{s} g =\\ \nonumber &&\int_{\Omega_{0}^{+}} d \hat{s} f + \int_{ \Omega_{+} } d \hat{s} g  =  \max_{{\cal A}} G(\rho;\hat{\beta};{\cal A},{\cal \tilde{T}}).      
\end{eqnarray}

As a result, for  $\alpha' > \alpha$ it holds that $\max_{{\cal A}} G(\rho;\hat{\beta'};{\cal A},{\cal \tilde{T}}) \geq \max_{{\cal A}} G(\rho;\hat{\beta};{\cal A},{\cal \tilde{T}})$.

{\it Proof of Fact 2 .- } 
It follows from the {\it Fact 1} that $max_{{\cal A}} G(\rho(\vec{x},\vec{y},-\lambda I);\hat{\beta};{\cal A},{\cal \tilde{T}})$ is monotonic in $\alpha$ parameter, where $\alpha = \left| \vec{y} \hat{u} \right|$. Let us consider a simple

{\it Lemma .- }
Let $f(a,x)$ will be a function with $a \in [a_0,a_1]$
and $x\in \Omega \subset R^{n}$ where $\Omega$ is compact.
Suppose that
(i) for any $a \leq a'$ and for any $x$ one has $f(a,x)\leq f(a',x)$;
(ii) the $x(a)$ is some (may be not unique) point realising maximum
of $f(a,x)$ over x for fixed $a$, i.e. $f(a,x(a))=max_x f(a,x)$.
Then the function $f(a,x(a))$ is monotonic in $a$. As a result $min_{a \in \left[a_0,a_1\right]} max_{x\in \Omega} f(a,x) = max_{x\in \Omega} f(a_0,x) $

{\it Proof of Lemma .- } Consider any $a \leq a'$.
Then we have $f(a,x(a))\leq f(a',x(a)) \leq f(a',x(a'))$, where the first inequality follows from (i) and the second one from (ii). 

Coming back to the proof of the Fact 2 we may put in place of $a \in \left[a_0,a_1\right]$ in the Lemma above 
the parameter $\alpha \in \left[0,1\right]$ and in place of $x$ all the other parameters contained in the sets ${\cal A},{\cal \tilde{T}}$
getting the desired monotonicity in Fact 2. 

 As a result of the Fact 2 we can set $\alpha = 0$ which implies $\hat{\beta}^* = \hat{y}$ $\Omega^+_0 = (0,\pi)$, $\Omega_+=\emptyset$ (what corresponds to von Neumann measurement of Alice). 
\begin{eqnarray}
&&min_{ \hat{\beta}} max_{{\cal A},{\cal \tilde{T}}} G(\rho(\vec{x},\vec{y},-\lambda I);\hat{\beta};{\cal A},{\cal \tilde{T}})= \\ \nonumber &&min_{ \hat{\beta}} max_{{\cal \tilde{T}}} G(\rho(\vec{x},\vec{y},-\lambda I);\hat{\beta};{\cal A^*},{\cal \tilde{T}}) =
\\ \nonumber &&  max_{{\cal \tilde{T}}} G(\rho(\vec{x},\vec{y},-\lambda I);\hat{\beta}^*=\hat{y};{\cal A^*},{\cal \tilde{T}}) = \\ \nonumber &&  max_{ \tilde
{T}^{(1)}_+, \tilde{T}^{(1)}_- } \int_{\Omega^+_0} d \hat{s} \lambda||(\tilde{T}^{(1)}_+-\tilde{T}^{(1)}_-)^{T}\hat{s}|| = \lambda 
\end{eqnarray} 
since (i) $\alpha=0$ implies $\vec{y} || \hat{\beta}$, (ii) the triangle inequality $||(\tilde{T}^{(1)}_+-\tilde{T}^{(1)}_-)^{T}\hat{s}|| \leq ||(\tilde{T}^{(1)}_+\hat{s}|| +||\tilde{T}^{(1)}_-\hat{s}|| \leq 2 $ is saturated  for the $\tilde{T}^{(1)}_{\pm}= \pm I$ choice and (iii) $d \hat{s}$ represents the measure on the plane $d \hat{s} = \frac{d \varphi}{2 \pi}$. Alice measurement is determined by $\hat{a}=\frac{M^{T}\hat{s}}{||M^{T}\hat{s}||}$. The choice of $\tilde{T}^{(1)}_{\pm}= \pm I$  implies that $M=2 \lambda I$ acts on the circle which eventually determines the Alice von Neumann measurement $\hat{a}=\hat{s}$. The latter together with $\tilde{T}^{(1)}_{\pm}= \pm I$ shows that the protocol optimal under quadratic fidelity in \cite{RSP} is also optimal here.

{\it Derivation of formula (12) (main text) for Bell diagonal states and bistochastic decodings.- }
Maximisation $\left\|M^T \hat{s}\right\|$ over bistochastic decoding strategies $\cal{\check{T}}$ gives $\check{T}_+ = I$ and $\check{T}_- = -I$ - rotation about $\hat{\beta}$ direction. Is not difficult to see, that as in the case of $LOCC^{ \rightarrow, 1}$ the optimal von Neumann measurement is determined by $\hat{a}=\hat{s}$. Again, for fixed $\hat{\beta}$, the protocol from \cite{RSP} turns out to be optimal. Now we should show that the minimisation of the protocol fidelity over $\hat{\beta}$ is provided for circle $\Omega$, which contains all versors $\hat{s}$ orthogonal to the eigenvector corresponding to the largest eigenvalue of $TT^T$. This means that RSP of pure states with Bloch vectors $\hat{s}$ from that circle is the least convenient from the point of view of transfer fidelity. For Bell states $TT^T = \text{diag}[t^2_1,t^2_2,t^2_3], \; t^2_3 \geq t^2_2 \geq t^2_1$. Let us parametrise all $\hat{s} \in \Omega $ (and hence orthogonal to the eigenvector corresponding to the largest eigenvalue of $TT^T$) by  $\varphi$ angle. Let us denote $max_{{\cal A}, {\cal \check{T}}} G(\rho(0,0,T);\hat{\beta};{\cal A},{\cal \check{T}})=\frac{1}{2 \pi}\int^{2 \pi}_{0} d \varphi g(\hat{s}(\varphi, \theta(\varphi)))$, where in this case $\theta(\varphi)$ is trivial ie. $\theta(\varphi)=0$ . In this setting we have we have   
\begin{eqnarray}
\label{LI}
&&\max_{{\cal A}, {\cal \check{T}}} G(\rho(0,0,T);\hat{\beta};{\cal A},{\cal \check{T}}) =   \\ \nonumber &&  \int d \hat{s} \left\|\left[(\check{T}_+ -\check{T}_-)T\right]^T \hat{s}\right\| = 2 \int d \hat{s}  \left\|T^T \hat{s}\right\| = \\ \nonumber  &&2 \int d \hat{s} \sqrt{\left\langle \hat{s}\right| T T^T \left| \hat{s} \right\rangle}  = \\ \nonumber  &&\frac{1}{2 \pi}\int^{2 \pi}_{0} d \varphi g(\hat{s}(\varphi, 0)) =  \\ \nonumber  &&\frac{1}{\pi} \int^{ \pi}_{0} d \varphi \sqrt{t^2_1 \cos^2 \varphi +t^2_2 \sin^2 \varphi }. 
\end{eqnarray}

\begin{figure}
	\centering
		\includegraphics[width=55mm]{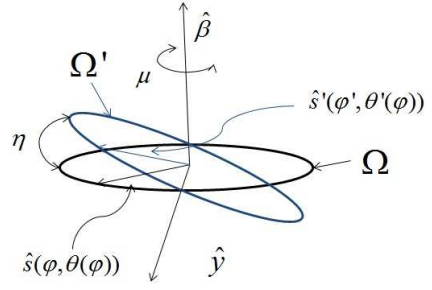}
	\caption{Relation between two coordinates systems. Plane containing circle $\Omega$ is orthogonal to $\hat{\beta}$. In this plane the unit versor $\hat{s}$ is parametrised be $\varphi$ angle and $g(\hat{s}(\varphi,0)))= \sqrt{\left\langle \hat{s}(\varphi,0)\right| T T^T \left| \hat{s}(\varphi,0) \right\rangle}$ . Plane containing $\Omega'$ is rotated with respect to that containing $\Omega$. Versor $\hat{s}'$ in the plane containing  $\Omega'$ can be again parametrised by $\varphi$ and we have $g(\hat{s}'(\varphi'(\varphi)))= \sqrt{\left\langle R_{\beta}(\mu)R_y(\eta) \hat{s}(\varphi,0)\right| T T^T \left| R_{\beta}(\mu)R_y(\eta) \hat{s}(\varphi,0) \right\rangle}$.}
\label{figsm}		
\end{figure}

Let us now consider a rotation of $\hat{s}$. We can decompose any rotation into rotation about y axis in a plane perpendicular to $\hat{\beta}$ by $\eta$ angle followed by rotation about $\hat{\beta}$ by $\mu$ angle. Then $\hat{s}$ is transformed into $\hat{s}'$, what corresponds to the change of parametrization $(\varphi, \theta(\varphi) )\rightarrow ( \varphi'(\varphi), \theta'(\varphi) )$ (see Fig. \ref{figsm}). In the rotated frame 
$max_{{\cal A}, {\cal T}} G(\rho(0,0,T);\hat{\beta'};{\cal A},{\cal \check{T}})=\int d \hat{s}' g(\hat{s}') = \frac{1}{2 \pi} \int^{2 \pi}_{0} d \varphi g(\hat{s}'(\varphi'(\varphi),\theta'(\varphi)))$ (for the explicit form of $g(\hat{s}'(\phi'(\varphi),\theta'(\varphi)))$ see (\ref{eq-g}) below).

We will need the following

{\it Lemma.- } The function $f(x) = \int^{2\pi}_{0} d \varphi \sqrt{A - B \sin^2 \varphi + x \sin 2 \varphi  }$ is decreasing function of x.

 {\it Proof.- }  We have 
\begin{eqnarray}
&& \frac{\partial}{\partial x} f(x) = \int^{2\pi}_0 d \varphi \frac{\sin \varphi \cos \varphi}{\sqrt{A - B \sin^2 \varphi + x \sin 2 \varphi}}  =
\\ \nonumber   &&2\int^{\frac{\pi}{2}}_0 d \varphi  \sin \varphi \cos \varphi  \left( \frac{1}{\sqrt{A - B \sin^2 \varphi + x \sin 2 \varphi}}\right. - \\ \nonumber &&\left. \frac{1}{\sqrt{A - B \sin^2 \varphi - x \sin 2 \varphi}} \right) < 0.
\end{eqnarray} 

\begin{widetext} 
To optimise the formula  (12) (main text),
we consider the function $g(\hat{s}'(\varphi'(\varphi),\theta'(\varphi)))$ which can be explicitly written as
\begin{eqnarray}
\label{eq-g}
&&g(\hat{s}'(\varphi'(\phi),\theta'(\varphi)))= \sqrt{\left\langle R_{\beta}(\mu)R_y(\eta) \hat{s}\right| T T^T \left| R_{\beta}(\mu)R_y(\eta) \hat{s} \right\rangle} =
 \\ \nonumber && \sqrt{ \cos^2 \varphi \left[\cos^2 \eta(t^2_1 \cos^2 \mu + t^2_2 \sin^2 \mu) +\sin^2 \eta t^2_3\right]+ \frac{1}{2} \sin 2 \varphi \sin 2 \mu \cos \eta (t^2_2-t^2_1) + \sin^2 \varphi (\sin^2 \mu t^2_1 + \cos^2 \mu t^2_2)}. 
\end{eqnarray}
Now the following relation holds
\begin{eqnarray}
&&\int^{2 \pi}_{0} d \varphi g(\hat{s}'(\varphi'(\phi),\theta'(\varphi))) = \\
\nonumber && \int^{2 \pi}_0 d \varphi \sqrt{ \cos^2 \varphi \left[\cos^2 \eta(t^2_1 \cos^2 \mu + t^2_2 \sin^2 \mu) +\sin^2 \eta t^2_3\right]+ \frac{1}{2} \sin 2 \varphi \sin 2 \mu \cos \eta (t^2_2-t^2_1) + \sin^2 \varphi (\sin^2 \mu t^2_1 + \cos^2 \mu t^2_2)} = \\ \nonumber && \int^{2 \pi}_0 d \varphi \sqrt{ (1-\sin^2 \varphi) \left[\cos^2 \eta(t^2_1 \cos^2 \mu + t^2_2 \sin^2 \mu) +\sin^2 \eta t^2_3\right]+ \frac{1}{2} \sin 2 \varphi \sin 2 \mu \cos \eta (t^2_2-t^2_1) + \sin^2 \varphi (\sin^2 \mu t^2_1 + \cos^2 \mu t^2_2)} \geq \\ \nonumber && \int^{2 \pi}_0 d \varphi \sqrt{\cos^2 \varphi \left[\cos^2 \eta(t^2_1 \cos^2 \mu + t^2_2 \sin^2 \mu) +\sin^2 \eta t^2_3\right]+ \frac{1}{2} \sin 2 \varphi \sin 2 \mu (t^2_2-t^2_1) + \sin^2 \varphi (\sin^2 \mu t^2_1 + \cos^2 \mu t^2_2)} \geq \\ \nonumber && \int^{2 \pi}_0 d \varphi \sqrt{\cos^2 \varphi (t^2_1 \cos^2 \mu + t^2_2 \sin^2 \mu)+ \frac{1}{2} \sin 2 \varphi \sin 2 \mu (t^2_2-t^2_1) + \sin^2 \varphi (\sin^2 \mu t^2_1 + \cos^2 \mu t^2_2)} = \\ \nonumber && \int^{2 \pi}_0 d \varphi \sqrt{t^2_1 \cos^2 \varphi +t^2_2 \sin^2 \varphi} = \\ \nonumber &&\int^{2 \pi}_{0} d \varphi g(\hat{s}(\varphi,0)). 
\end{eqnarray}
The first inequality follows from Lemma with $A= \cos^2 \eta(t^2_1 \cos^2 \mu + t^2_2 \sin^2 \mu) +\sin^2 \eta t^2_3$, $B= \cos^2 \eta(t^2_1 \cos^2 \mu + t^2_2 \sin^2 \mu) +\sin^2 \eta t^2_3 - \sin^2 \mu t^2_1 + \cos^2 \mu t^2_2$ and $x=\frac{1}{2} \sin 2 \varphi \sin 2 \mu \cos \eta (t^2_2-t^2_1)$. We increased $x$ by setting $\cos \eta = 1$. The second inequality follows from the fact that $t^2_3\geq \sin^2 \mu t^2_1 + \cos^2 \mu t^2_2$, the last equality from the fact that rotation about $\hat{\beta}$ direction by the angle $\mu$ does not change the value of the function.   
\end{widetext}
Because $ \int^{2 \pi}_{0} d \phi g(\hat{s}(\varphi,0)) \leq \int^{2 \pi}_{0} d \phi g(\hat{s}'(\varphi',\theta'(\varphi)))$, 
minimum over $\hat{\beta}$ provides $\hat{s}$ orthogonal to the largest eigenvalue of $TT^T$. Taking this into account we have   
\begin{eqnarray}
\label{LI}
&&\min_{\beta}\max_{{\cal A}, {\cal \check{T}}} G(\rho(0,0,T);\hat{\beta};{\cal A},{\cal \check{T}}) = \\ \nonumber && \frac{1}{2 \pi}\int^{2 \pi}_{0} d \phi g(\hat{s}(\varphi,0)) \\ \nonumber  &&\frac{1}{\pi} \int^{ \pi}_{0} d \varphi \sqrt{t^2_1 \cos^2 \varphi +t^2_2 \sin^2 \varphi }
\end{eqnarray}
Let us recall definition of the complete elliptic integral of the second kind \cite{NIST}
\begin{equation}
\begin{split}
E(k) = \int^{\frac{\pi}{2}}_{0}  d \varphi  \sqrt{1-k^2 \sin^2 \varphi}.
\end{split}
\end{equation}
Using this we get 
\begin{eqnarray}
&& \min_{\beta} \max_{{\cal A},{\cal \check{T}}} G(\rho(0,0,T);\hat{\beta};{\cal A},{\cal \check{T}}) = \nonumber \\
&&=\frac{2 |t_2|}{\pi} E \left(\sqrt{1-\frac{t^2_1}{t^2_2}}\right).
\end{eqnarray}


\end{document}